\documentclass[preprint]{aastex}

\begin{document}

\title{An Analytic Theory for the Orbits of Circumbinary Planets}

\author{Gene C. K. Leung\altaffilmark{1} and
        Man Hoi Lee\altaffilmark{1,2}}
\altaffiltext{1}{Department of Physics, The University of Hong Kong,
  Pokfulam Road, Hong Kong}
\altaffiltext{2}{Department of Earth Sciences, The University of Hong Kong,
  Pokfulam Road, Hong Kong}

\begin{abstract}
Three transiting circumbinary planets (Kepler-16~b, Kepler-34~b, and
Kepler-35~b) have recently been discovered from photometric data taken
by the {\it Kepler} spacecraft.
Their orbits are significantly non-Keplerian because of the large
secondary-to-primary mass ratio and orbital eccentricity of the
binaries, as well as the proximity of the planets to the binaries.
We present an analytic theory, with the planet treated as a test
particle, which shows that the planetary motion can be represented by
the superposition of the circular motion of a guiding center, the
forced oscillations due to the non-axisymmetric components of the
binary's potential, the epicyclic motion, and the vertical motion.
In this analytic theory, the periapse and ascending node of the planet
precess at nearly equal rates in opposite directions.
The largest forced oscillation term corresponds to a forced
eccentricity (which is an explicit function of the parameters of the
binary and of the guiding center radius of the planet), and the
amplitude of the epicyclic motion (which is a free parameter of the
theory) is the free eccentricity.
Comparisons with direct numerical orbit integrations show that this
analytic theory gives an accurate description of the planetary motion
for all three Kepler systems.
We find that all three Kepler circumbinary planets have nonzero free
eccentricities.
\end{abstract}

\section{INTRODUCTION}

\citet{doy11} have recently discovered the first transiting
circumbinary planet, Kepler-16~b, from photometric data taken by
the {\it Kepler} spacecraft.
The Saturn-mass planet orbits a pair of stars of $0.69 M_\sun$ and
$0.20 M_\sun$.
\cite{wel12} subsequently announced the discovery of two more
circumbinary planets: Kepler-34~b and Kepler-35~b.
Kepler-34~b is $0.22 M_{\rm J}$ (where $M_{\rm J}$ is the mass of
Jupiter) and orbits two Sun-like stars, while Kepler-35~b is $0.13
M_{\rm J}$ and orbits a pair of smaller stars ($0.89M_\sun$ and $0.81
M_\sun$).
For all three systems, the orbits of the binary and planet are aligned
to within $2^\circ$.
From the observed rate of circumbinary planets in their sample,
\cite{wel12} estimated that more than $\sim 1\%$ of close binary stars
have giant planets on nearly coplanar orbits.

Variations in eclipse times and transit durations, combined with
radial velocity measurements, allow precise measurements of both
physical and orbital parameters for all three systems.
Table \ref{table1} shows the best-fit osculating Keplerian orbital
parameters provided by J.~A.~Carter (2012, private communication).
They differ slightly from those published in Table 1 of \cite{wel12},
as the values in that table are the medians of the cumulative
distribution of the marginalized posteriors for each parameter, and
they are osculating parameters at a different epoch.

Plots of the evolution of the osculating Keplerian orbital elements of
the planets in the Supporting Online Material of \cite{doy11} and
Supplementary Information of \cite{wel12} show significant variations
on both orbital and secular timescales, with the eccentricity changing
from nearly zero to 0.1 in the case of Kepler-16~b, and the precession
period of the orbit is as short as $\sim 60$ orbital periods in the
case of Kepler-35~b (see figures below for more details).
The nontrivial departures from unperturbed Keplerian orbits for these
circumbinary planets are due to the large secondary-to-primary stellar
mass ratio ($m_B/m_A = 0.29$--$0.97$), the large orbital eccentricity
of the binary $e_{AB} = 0.14$--$0.52$, and the proximity of the planet
to the binary (orbital period ratio $P_b/P_{AB} = 5.6$--$10.4$).

In \S \ref{sec:theory} we present an analytic theory for the
orbit of a circumbinary planet that is valid in the limit that the
planet has negligible mass and can be treated as a test particle.
\cite{lee06} have previously developed an analytic theory for the
orbits of the small satellites of the Pluto-Charon system, assuming
that the orbit of Charon relative to Pluto is circular.
We generalize that theory to the case of an eccentric binary orbit.
In \S \ref{sec:numerical} we present the results of direct
numerical orbit integrations and compare them with the analytic
theory.
In \S \ref{sec:discussion} we discuss the limitations of the analytic
theory and a simple modification that can significantly improve the
analytic predictions when the orbital eccentricity of the binary is
large.
Our conclusions are summarized in \S \ref{sec:summary}.

\section{ANALYTIC THEORY}
\label{sec:theory}

In this section we follow a similar approach as \citet{lee06} to
develop an analytic theory for the orbit of a circumbinary planet.
We extend their theory to first order in the eccentricity of the orbit
of the binary.
We assume that the planet has negligible mass and can be treated as a
test particle.
Then the orbit of the secondary (hereafter B) relative to the primary
(hereafter A) is an elliptical Keplerian orbit with eccentricity
$e_{AB}$ and semimajor axis $a_{AB}$, and the distance between A and B
is $r_{AB} = a_{AB} (1 - e_{AB}^2)/(1 + e_{AB}\cos f_B)$, where $f_B$
is the true anomaly of B.
We adopt a cylindrical coordinate system with the origin at the center
of mass of the binary and the $x$-$y$ plane being the orbital plane of
the binary.
The positions of B and A are $\mbox{\boldmath $r$}_B = (R_B, \phi_B,
0)$ and $\mbox{\boldmath $r$}_A = (R_A, \phi_B+\pi, 0)$, respectively,
where $R_B = r_{AB} m_A/(m_A+m_B)$, $R_A = r_{AB} m_B/(m_A+m_B)$,
$m_A$ is the mass of A, and $m_B$ is the mass of B.

\subsection{Potential}
\label{sec:potential}

The gravitational potential at $\mbox{\boldmath $r$} = (R, \phi, z)$
due to the binary is
\begin{equation}
\Phi(\mbox{\boldmath $r$}) =
- {G m_A \over \left|\mbox{\boldmath $r$} - \mbox{\boldmath $r$}_A\right|}
- {G m_B \over \left|\mbox{\boldmath $r$} - \mbox{\boldmath $r$}_B\right|} .
\end{equation}
Since the orbit of the planet is nearly coplanar with that of the
binary, we expand $1/|\mbox{\boldmath $r$} - \mbox{\boldmath $r$}_B|$
in powers of $z$:
\begin{equation}
{1 \over \left|\mbox{\boldmath $r$} - \mbox{\boldmath $r$}_B\right|}
= {1 \over \left(\rho^2 + z^2\right)^{1/2}}
= {1 \over \rho} - {1 \over 2} {z^2 \over \rho^3} + \cdots ,
\end{equation}
where
\begin{equation}
\rho = \left[R^2 + R_B^2 - 2 R R_B \cos\left(\phi-\phi_B\right)\right]^{1/2} .
\end{equation}
The inverse powers of $\rho$ can be expressed as cosine series using
equation (6.62) of \cite{mur99} to give
\begin{equation}
{1 \over \left|\mbox{\boldmath $r$} - \mbox{\boldmath $r$}_B\right|}
= {1 \over 2 R} \sum_{k=0}^{\infty} (2-\delta_{k0})
  \left[b_{1/2}^k(R_B/R) - {1 \over 2} \left(z \over R\right)^2
        b_{3/2}^k(R_B/R) + \cdots\right] \cos k(\phi-\phi_B) ,
\end{equation}
where $\delta_{k0}$ is the Kronecker delta and $b_s^k(R_B/R)$ are the
Laplace coefficients.
Similarly,
\begin{equation}
{1 \over \left|\mbox{\boldmath $r$} - \mbox{\boldmath $r$}_A\right|}
= {1 \over 2 R} \sum_{k=0}^{\infty} (-1)^k (2-\delta_{k0})
  \left[b_{1/2}^k(R_A/R) - {1 \over 2} \left(z \over R\right)^2
        b_{3/2}^k(R_A/R) + \cdots\right] \cos k(\phi-\phi_B) .
\end{equation}
Then the potential can be written as
\begin{equation}
\Phi(\mbox{\boldmath $r$}) =
  \sum_{k=0}^{\infty} \left[
  \Phi_{0k} (R) - {1 \over 2} \left(z \over R\right)^2 \Phi_{2k} (R) +
  \cdots \right] \cos k(\phi-\phi_B) ,
\end{equation}
where
\begin{eqnarray}
\Phi_{jk} (R) &=& - {2-\delta_{k0} \over 2} \Bigg[
                (-1)^k{m_A \over (m_A+m_B)} b_{(j+1)/2}^k(R_A/R) \nonumber \\
&&\qquad + {m_B \over (m_A+m_B)} b_{(j+1)/2}^k(R_B/R)\Bigg]
                 {G (m_A+m_B) \over R} .
\end{eqnarray}

To first order in $e_{AB}$,
\begin{equation}
\phi_B = f_B + \varpi_B \approx M_B + 2 e_{AB} \sin M_B + \varpi_B ,
\end{equation}
where $\varpi_B$ and $M_B = n_{AB} t+ \varphi_{AB}$ are, respectively,
the longitude of periapse and mean anomaly of B relative A, $n_{AB} =
[G (m_A + m_B)/a_{AB}^3]^{1/2}$ is the mean motion of the binary, and
$\varphi_{AB}$ is a constant.
Then
\begin{eqnarray}
\cos k(\phi-\phi_B) & \approx & \cos k(\phi-M_B-\varpi_B) 
+ e_{AB} k \lbrack\cos (k(\phi-\varpi_B)-(k+1) M_B) \nonumber \\ 
& & -\cos (k(\phi-\varpi_B)-(k-1) M_B)\rbrack .
\end{eqnarray}
To first order in $e_{AB}$,
\begin{equation}
R_B/a_B \approx 1 - e_{AB} \cos M_B ,
\end{equation}
and
\begin{equation}
b_{(j+1)/2}^k(R_B/R) \approx b_{(j+1)/2}^k(\alpha_{B}) - e_{AB}
\alpha_{B} Db_{(j+1)/2}^k(\alpha_{B})\cos M_B ,
\label{lap}
\end{equation}
where $a_B = a_{AB} m_A/(m_A + m_B)$, $\alpha_{B} = a_B/R$, and $D =
d/d\alpha$.
Equation (\ref{lap}) gives rise to terms involving 
\begin{equation}
e_{AB} \cos M_B \cos k(\phi-\phi_B)={e_{AB}\over 2}\left[\cos (k(\phi-\phi_B) + M_B) + \cos (k(\phi-\phi_B)- M_B)\right],
\end{equation}
which can be expressed as
\begin{equation}
e_{AB} \cos (k(\phi-\phi_B)\pm M_B) \approx e_{AB} \cos (k(\phi-\varpi_B)-(k\mp 1) M_B).
\end{equation}
Terms involving $R_A/R$ can be found in a similar manner.

After re-grouping the terms,
\begin{eqnarray}
\Phi(\mbox{\boldmath $r$}) &=& \sum_{k=0}^{\infty} \left[ \Phi_{0k0}(R)
- {1 \over 2} \left(z \over R\right)^2 \Phi_{2k0} (R) + \cdots
  \right] \cos k(\phi-M_B-\varpi_B)
\label{Phi} \\ 
&&+ e_{AB} \sum_{k=0}^{\infty} \Bigg\{ k\left[\Phi_{0k0} (R) - {1 \over 2}
  \left(z \over R\right)^2 \Phi_{2k0} (R) + \cdots\right] \nonumber \\
&&\qquad - {1 \over 2} \left[\Phi_{0k1} (R) - {1 \over 2} \left(z \over R\right)^2 \Phi_{2k1} (R) + \cdots\right]\Bigg\} \cos (k(\phi-\varpi_B)-(k+1) M_B) \nonumber \\ 
&&+ e_{AB} \sum_{k=0}^{\infty} \Bigg\{-k\left[\Phi_{0k0} (R) - {1 \over 2}
  \left(z \over R\right)^2 \Phi_{2k0} (R) + \cdots\right] \nonumber \\
&&\qquad - {1 \over 2} \left[\Phi_{0k1} (R) - {1 \over 2} \left(z
    \over R\right)^2 \Phi_{2k1} (R) + \cdots\right]\Bigg\} \cos
(k(\phi-\varpi_B)-(k-1) M_B), \nonumber
\end{eqnarray}
where
\begin{eqnarray}
\Phi_{jk0} (R) &=& - {2-\delta_{k0} \over 2} \Bigg[
                (-1)^k{m_A \over (m_A+m_B)} b_{(j+1)/2}^k(\alpha_{A})\nonumber\\
&&\qquad +  {m_B \over (m_A+m_B)} b_{(j+1)/2}^k(\alpha_{B})
                \Bigg] {G (m_A+m_B) \over R} 
\label{Phijk0}
\end{eqnarray}
and
\begin{eqnarray}
\Phi_{jk1} (R) &=& - {2-\delta_{k0} \over 2} \Bigg[
                (-1)^k{m_A \over (m_A+m_B)}
                \alpha_{A}Db_{(j+1)/2}^k(\alpha_{A})\nonumber\\
&&\qquad + {m_B \over (m_A+m_B)} \alpha_{B}Db_{(j+1)/2}^k(\alpha_{B})
                \Bigg] {G (m_A+m_B) \over R} .
\label{Phijk1}
\end{eqnarray}
The terms in equation (\ref{Phi}) multiplied by $e_{AB}$ are new
compared to the potential due to a binary on circular orbit derived by
\cite{lee06}.
As in the circular binary orbit case, the axisymmetric $\Phi_{j00}$
components of the potential are identical to those due to two rings:
one of mass $m_A$ and radius $a_A$ and another of mass $m_B$ and
radius $a_B$.

\subsection{Equations of Motion and Solution}
\label{sec:solution}

The equations of motion in cylindrical coordinates are
\begin{eqnarray}
\ddot{R} - R {\dot{\phi}}^2 &=&
     - {\partial \Phi \over \partial R} , \nonumber \\
R \ddot{\phi} + 2 \dot{R} \dot{\phi} &=&
     - {1 \over R}{\partial \Phi \over \partial \phi} , \label{EOM} \\
\ddot{z} &=&
     - {\partial \Phi \over \partial z} . \nonumber
\end{eqnarray}
As in \cite{lee06}, we approximate the orbit of the circumbinary
planet as small deviations from the circular motion of a guiding
center in the $x$-$y$ plane:
\begin{eqnarray}
R &=& R_0 + R_1(t) , \nonumber \\
\phi &=& \phi_0(t) + \phi_1(t) , \label{Rphiz} \\
z &=& z_1(t) , \nonumber
\end{eqnarray}
where the constant $R_0$ is the radius of the guiding center,
$|R_1/R_0| \ll 1$, $|\phi_1| \ll 1$, and $|z_1/R_0| \ll 1$.

Substituting equations (\ref{Phi}) and (\ref{Rphiz}) into equation
(\ref{EOM}), the only nontrivial equation at the zeroth order is
\begin{equation}
R_0 {\dot{\phi_0}}^2 = \left[d\Phi_{000} \over dR\right]_{R_0} ,
\label{dotphi0}
\end{equation}
which describes the circular motion of the guiding center.
Its solution is
\begin{equation}
\phi_0(t) = n_0 t + \varphi_0 ,
\end{equation}
where $\varphi_0$ is a constant and the mean motion $n_0$ is given by
\begin{eqnarray}
n_0^2 
&=& \left[{1 \over R}{d\Phi_{000} \over dR}\right]_{R_0} \label{n0} \\
&=& {1 \over 2} \Bigg\{
    {m_A \over (m_A+m_B)} b_{1/2}^0(\alpha_{A}) +
    {m_B \over (m_A+m_B)} b_{1/2}^0(\alpha_{B}) \nonumber \\
& & \qquad + {m_A m_B \over (m_A + m_B)^2} \left(a_{AB} \over R_0\right)
             \left[Db_{1/2}^0(\alpha_{A}) + Db_{1/2}^0(\alpha_{B})
             \right]
\Bigg\} n_K^2 . \nonumber
\end{eqnarray}
In the above equation $n_K = [G (m_A+m_B)/R_0^3]^{1/2}$ is the
Keplerian mean motion at $R_0$, and $\alpha_{A}$ and
$\alpha_{B}$ are evaluated at $R = R_0$.

At the first order, the equations of motion are
\begin{eqnarray}
\ddot{R}_1 &=& 2 R_0 n_0 \dot{\phi}_1 - \left[ {d^2\Phi_{000} \over dR^2} - n^2 \right]_{R_0} R_1 + e_{AB} \left[{d\Phi_{001} \over dR}\right]_{R_0} \cos M_B
\nonumber\\
& &  - \sum_{k=1}^{\infty}\Bigg\{\left[{d\Phi_{0k0} \over dR}\right]_{R_0} \cos k(\phi_0-M_B-\varpi_B)\nonumber\\
& & -e_{AB} \left[-k {d\Phi_{0k0} \over dR} + {1\over 2} {d\Phi_{0k1} \over dR}\right]_{R_0} \cos (k(\phi_0-\varpi_B)-(k+1)M_B)\nonumber\\
& & -e_{AB} \left[k {d\Phi_{0k0} \over dR} + {1\over 2} {d\Phi_{0k1} \over dR}\right]_{R_0} \cos (k(\phi_0-\varpi_B)-(k-1)M_B)\Bigg\} ,
\label{ddotr11}\\
\ddot{\phi}_1 &=& -{2 n_0 \over R_0} \dot{R}_1 
+ \sum_{k=1}^{\infty} {k\over R_0^2} \Bigg\{\Phi_{0k0}(R_0)\sin k(\phi_0-M_B-\varpi_B)\nonumber\\
& & - e_{AB} \left[-k \Phi_{0k0} + {1\over 2}\Phi_{0k1}\right]_{R_0} \sin (k(\phi_0-\varpi_B)-(k+1)M_B)\nonumber\\
& & - e_{AB} \left[k \Phi_{0k0} + {1\over 2}\Phi_{0k1}\right]_{R_0} \sin (k(\phi_0-\varpi_B)-(k-1)M_B) \Bigg\}  ,
\label{ddotphi1}\\
\ddot{z}_1 &=& \left[\Phi_{200} \over R^2\right]_{R_0} z_1 ,
\label{ddotz1}
\end{eqnarray}
where $n = (R^{-1} d\Phi_{000}/dR)^{1/2}$ is the mean motion at $R$,
and the quantities in the square brackets with the subscript $R_0$ are
evaluated at $R = R_0$.
Equation (\ref{ddotphi1}) can be integrated to give $\dot{\phi}_1$,
which can then be substituted into equation (\ref{ddotr11}) to yield
\begin{eqnarray}
\ddot{R}_1 + \kappa_0^2 R_1 &=& e_{AB} \left[{d\Phi_{001} \over dR}\right]_{R_0} \cos M_B \nonumber\\
&&- \sum_{k=1}^{\infty} \Bigg\{ \left[{d\Phi_{0k0} \over dR} + {2 n \Phi_{0k0} \over R (n - n_{AB})}
     \right]_{R_0} \cos k(\phi_0-M_B-\varpi_B)\nonumber\\ 
&&- e_{AB} \left[-k {d\Phi_{0k0} \over dR}
 + {1\over 2} {d\Phi_{0k1} \over dR}
 + {k n (-2k\Phi_{0k0} + \Phi_{0k1})\over R (k n - (k+1)n_{AB})} \right]_{R_0} \nonumber\\
&&\times \cos (k(\phi_0-\varpi_B)-(k+1)M_B) \nonumber\\ 
&&- e_{AB} \left[ k {d\Phi_{0k0} \over dR}
 + {1\over 2} {d\Phi_{0k1} \over dR}
 + {k n (2k\Phi_{0k0} + \Phi_{0k1})\over R (k n - (k-1)n_{AB})} \right]_{R_0} \nonumber\\
&&\times  \cos (k(\phi_0-\varpi_B)-(k-1)M_B)
\Bigg\} ,
\label{ddotR1}
\end{eqnarray}
where the epicyclic frequency $\kappa_0$ is given by
\begin{eqnarray}
\kappa_0^2
&=& \left[R{dn^2 \over dR} + 4n^2\right]_{R_0}
\label{kappa0} \\
&=& {1 \over 2} \Bigg\{
    {m_A \over (m_A+m_B)} b_{1/2}^0(\alpha_{A}) +
    {m_B \over (m_A+m_B)} b_{1/2}^0(\alpha_{B})
\nonumber \\
& & \qquad - {m_A m_B \over (m_A + m_B)^2} \left(a_{AB} \over R_0\right)
             \left[Db_{1/2}^0(\alpha_{A}) + Db_{1/2}^0(\alpha_{B})
             \right]
\nonumber \\
& & \qquad - {m_A m_B \over (m_A + m_B)^2} \left(a_{AB} \over R_0\right)^2
             \Bigg[{m_B \over (m_A+m_B)} D^2b_{1/2}^0(\alpha_{A})
\nonumber \\
& & \qquad         + {m_A \over (m_A+m_B)} D^2b_{1/2}^0(\alpha_{B})\Bigg]
\Bigg\} n_K^2 .
\nonumber
\end{eqnarray}
Equation (\ref{ddotR1}) is the equation of motion of a simple harmonic
oscillator of natural frequency $\kappa_0$ that is driven at
frequencies $n_{AB}$, $k|n_0-n_{AB}|$, and $|kn_0-(k\pm 1)n_{AB}|$,
and its solution gives
\begin{eqnarray}
R &=&
R_0 \Bigg\{1 - e_{\rm free} \cos(\kappa_0 t + \psi) - C_0 \cos M_B - \sum_{k=1}^{\infty} \Big[C_k^0 \cos k(\phi_0-M_B-\varpi_B) \nonumber\\
& & + C_k^+ \cos (k(\phi_0-\varpi_B)-(k+1)M_B)
+ C_k^- \cos (k(\phi_0-\varpi_B)-(k-1)M_B)
\Big]
\Bigg\} , \label{Rt} 
\end{eqnarray}
where $e_{\rm free}$ and $\psi$ are arbitrary constants and
\begin{eqnarray}
C_0 &=& - e_{AB} \left[{d\Phi_{001} \over dR}\right]_{R_0} \Bigg/
\left[R_0 (\kappa_0^2 - n_{AB}^2)\right] , \\
C_k^0 &=& \left[{d\Phi_{0k0} \over dR} + {2 n \Phi_{0k0} \over R (n - n_{AB})}
            \right]_{R_0} \Bigg/ \left\{R_0 \left[\kappa_0^2 -
                k^2(n_0-n_{AB})^2\right]\right\} ,
\label{Ck0} \\
C_k^\pm &=& e_{AB}\left[ \pm k{d\Phi_{0k0} \over dR}
 - {1\over 2} {d\Phi_{0k1} \over dR}
 + {k n (\pm 2 k\Phi_{0k0} - \Phi_{0k1}) \over R (k n -
   (k\pm 1)n_{AB})} \right]_{R_0} \Bigg/
\nonumber \\
& & \left\{R_0\left[\kappa_0^2-(k n_0 - (k\pm 1)
    n_{AB})^2\right]\right\} .
\label{Ckpm}
\end{eqnarray}
We can then solve equation (\ref{ddotphi1}) to give
\begin{eqnarray}
\phi &=&
 n_0 t + \varphi_0
+ {2 n_0 \over \kappa_0} e_{\rm free} \sin(\kappa_0 t + \psi)
+ {n_0 \over n_{AB}} D_0 \sin M_B
\nonumber\\
& & + \sum_{k=1}^{\infty} \Big[
{n_0 \over k (n_0 - n_{AB})} D_k^0 \sin k(\phi_0-M_B-\varpi_B)
\nonumber\\
& & + {n_0 \over k n_0 - (k+1) n_{AB}} D_k^+ \sin (k(\phi_0-\varpi_B)-(k+1)M_B)
\nonumber\\
& & + {n_0 \over k n_0 - (k-1) n_{AB}} D_k^- \sin (k(\phi_0-\varpi_B)-(k-1)M_B)
\Big] ,
\label{phit}
\end{eqnarray}
where
\begin{eqnarray}
D_0 &=& 2 C_0 , \\
D_k^0 &=& 2 C_k^0 -
\left[\Phi_{0k0} \over R^2 n (n - n_{AB})\right]_{R_0} ,
 \\
D_k^\pm &=& 2 C_k^\pm -
 e_{AB}\left[{k (\pm 2 k\Phi_{0k0} - \Phi_{0k1}) \over 2 R^2 n (k n -
   (k\pm 1)n_{AB})} \right]_{R_0} .
\end{eqnarray}

The motion in $R$ and $\phi$ is the superposition of the circular
motion of the guiding center at $R_0$ at frequency $n_0$, the
epicyclic motion represented by the {\it free} eccentricity $e_{\rm free}$
at frequency $\kappa_0$, and the forced oscillations of fractional radial
amplitudes $C_0$, $C_k^0$ and $C_k^\pm$ at frequencies $n_{AB}$,
$k|n_0-n_{AB}|$ and $|kn_0-(k\pm 1)n_{AB}|$, respectively.
Note that $C_0$, $C_k^0$, or $C_k^\pm$ is singular if
$\kappa_0^2 - n_{AB}^2$, $k n_0 - l n_{AB}$, or $\kappa_0^2 -
(k n_0 - l n_{AB})^2 = 0$, where $l = k$ or $k \pm 1$.
The second and third combinations of frequencies correspond to the
corotation and Lindblad resonances, respectively (e.g.,
\citealt{gol80}).
None of these resonances could be encountered if the planet is further
away than the 3:1 mean-motion resonance with the binary.

For the motion in $z$, the solution to equation (\ref{ddotz1}) is
\begin{equation}
z = z_1 = R_0 i_{\rm free} \cos(\nu_0 t + \zeta) ,
\end{equation}
where $i_{\rm free}$ and $\zeta$ are arbitrary constants and the
vertical frequency $\nu_0$ is given by
\begin{eqnarray}
\nu_0^2
&=& \left[-{\Phi_{200} \over R^2}\right]_{R_0}
\label{nu0} \\
&=& {1 \over 2} \left[
    {m_A \over (m_A+m_B)} b_{3/2}^0(\alpha_{A}) +
    {m_B \over (m_A+m_B)} b_{3/2}^0(\alpha_{B})
    \right] n_K^2 .
\nonumber
\end{eqnarray}
Thus the motion in $z$ decouples from that in $R$ and $\phi$ and has
only free oscillations represented by the {\it free} inclination
$i_{\rm free}$ at the vertical frequency $\nu_0$.

As we shall see, circumbinary planets have $\nu_0 > n_0 > n_K >
\kappa_0$.
So the azimuthal period $2\pi/n_0$ is shorter than the Keplerian
orbital period $2\pi/n_K$, the periapse precesses in the prograde
direction with the period $2\pi/|n_0 - \kappa_0|$, and the ascending
node precesses in the retrograde direction with the period $2\pi/|n_0
- \nu_0|$.

As in the circular binary orbit theory of \cite{lee06}, the motion is
represented by the circular motion of the guiding center, the
epicyclic motion, the forced oscillations and the vertical motion.
The expressions for the mean motion $n_0$ (eq.~[\ref{n0}]), the
epicyclic frequency $\kappa_0$ (eq.~[\ref{kappa0}]), and the vertical
frequency $\nu_0$ (eq.~[\ref{nu0}]) involve the axisymmetric
$\Phi_{000}$ and $\Phi_{200}$ components of the potential and are
identical to those in the circular binary orbit case
(there are, however, corrections at the second order in $e_{AB}$; see
\S \ref{sec:discussion}).
The motion in $z$ is identical to the circular binary orbit case.
The forced oscillations are composed of both terms
identical to those in the circular binary orbit theory ($C_k^0$ at
frequencies $k|n_0-n_{AB}|$) and new terms ($C_0$ at frequency
$n_{AB}$ and $C_k^\pm$ at frequencies $|kn_0-(k\pm 1)n_{AB}|$).
Note that the new terms $C_0$ and $C_k^\pm$ are proportional to
$e_{AB}$ and that one of these terms, $C_1^-$, has frequency $n_0$ and
can be identified as the {\it forced} eccentricity.
The {\it forced} longitude of periapse is aligned with the binary's
periapse because $C_1^- \ge 0$.

If we expand the analytic expressions in powers of $a_{AB}/R_0$ (note
that $a_{AB}/R_0 \la 0.32$ for the three Kepler systems) and retain
only the lowest order term, we find that the precession rates
\begin{equation}
{n_0-\kappa_0 \over n_K} \approx - {n_0-\nu_0 \over n_K} \approx
{3 \over 4} {m_A m_B \over (m_A + m_B)^2} \left(a_{AB} \over R_0\right)^2 ,
\label{preratesapprox}
\end{equation}
the forced eccentricity
\begin{equation}
C_1^- \approx {5 \over 4} e_{AB} {(m_A - m_B) \over (m_A + m_B)} \left(a_{AB} \over R_0\right) ,
\label{C1mapprox}
\end{equation}
and the other forced oscillation terms (up to $k = 3$)
\begin{eqnarray}
C_2^0 &\propto& {m_A m_B \over (m_A + m_B)^2} \left(a_{AB} \over
R_0\right)^5 , \\
C_0,\ C_2^\pm &\propto& e_{AB} {m_A m_B \over (m_A + m_B)^2}
\left(a_{AB} \over R_0\right)^5 , \\
C_1^0,\ C_3^0 &\propto& {m_A m_B (m_A - m_B) \over (m_A + m_B)^3}
\left(a_{AB} \over R_0\right)^6 ,
\label{C10C30approx} \\
C_1^+,\ C_3^\pm &\propto& e_{AB} {m_A m_B (m_A - m_B) \over (m_A +
  m_B)^3} \left(a_{AB} \over R_0\right)^6 .
\label{C1pC3pmapprox}
\end{eqnarray}
We can see from equations (\ref{Ck0}) and (\ref{Ckpm}) that $C_k^0$
and $C_k^\pm$ involve $\Phi_{0k0}$, $\Phi_{0k1}$, and their
derivatives with respect to $R$.
According to equations (\ref{Phijk0}) and (\ref{Phijk1}), these terms
would be exactly zero if $k$ is odd and $m_A = m_B$.
Equations (\ref{C1mapprox}), (\ref{C10C30approx}), and
(\ref{C1pC3pmapprox}) show that the odd terms are proportional to
$(m_A - m_B)/(m_A + m_B)$ at the lowest order in $a_{AB}/R_0$ and
could be small if $m_A \approx m_B$.

\section{COMPARISONS WITH NUMERICAL ORBIT INTEGRATIONS}
\label{sec:numerical}

For the numerical orbit integrations, we use Jacobi coordinates (where
the position of the secondary B is relative to the primary A and the
position of the planet is relative to the center of mass of the
binary), with the invariable plane as the reference plane.
This coordinate system reduces to that used in \S \ref{sec:theory}
when the mass of the planet is negligible.
We perform direct numerical orbit integrations of the Kepler-16, 34,
and 35 systems, using the \cite{wis91} symplectic integrator with the
modification described in \cite{lee03}.
The modification allows the integration of circumbinary planets
without an excessively small timestep, and we use a timestep of $0.1$
day.
We generate the initial positions and velocities of the binary and
planet by using the best-fit osculating orbital parameters in Table
\ref{table1}.

For comparison with the analytic theory, we need to evaluate $n_K$,
$n_0$, $\kappa_0$, $\nu_0$, and the fractional amplitudes $C_0$,
$C_k^0$, and $C_k^\pm$ at a guiding center radius $R_0$.
For each system, we adopt the average of the maximum and minimum
values of the cylindrical radius $R$ of the planet's orbit in the
numerical orbit integration over many precession cycles as $R_0$.
The adopted $R_0$ and the numerical values of $n_K$, $n_0/n_K$,
$\kappa_0/n_K$, $\nu_0/n_K$, $C_0$, $C_k^0$, and $C_k^\pm$ ($k = 1$,
$2$, and $3$) are listed in Table \ref{table2}.

\subsection{Kepler-16}
\label{sec:kepler16}

\begin{figure}
\epsscale{0.65}
\plotone{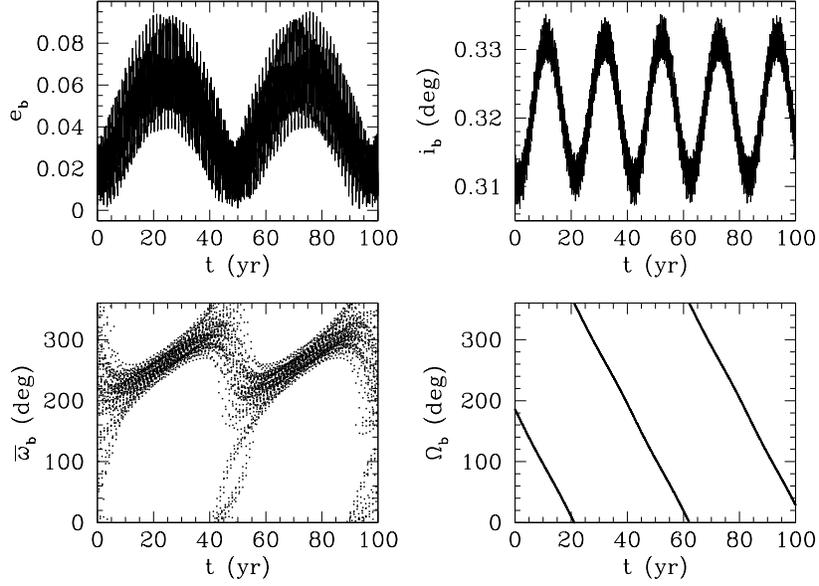}
\caption{
Evolution of the osculating Keplerian orbital elements (eccentricity
$e_b$, inclination $i_b$, longitude of periapse $\varpi_b$, and
longitude of the ascending node $\Omega_b$) of the planet Kepler-16~b
over $100\,$yr from direct numerical orbit integration.
The elements are relative to the center of mass of the binary, and the
reference plane is the invariable plane.
\label{fig:16orbel}
}
\end{figure}

In Figure \ref{fig:16orbel} we plot the evolution of the osculating
Keplerian orbital elements of the planet Kepler-16~b over $100$ years
from the numerical orbit integration.
The eccentricity $e_b$ shows variations on both orbital and apsidal
precession timescales.
The longitude of periapse $\varpi_b$ changes rapidly when $e_b$ is
nearly zero, but the long-term trend is prograde precession with
a period of $48.6$ years.
The precession of the longitude of ascending node $\Omega_b$ is
retrograde with a period of $41.0$ years, and the inclination $i_b$
shows small oscillations around a constant value, with two
oscillations per nodal precession period.
Using $n_K$, $n_0/n_K$, $\kappa_0/n_K$, and $\nu_0/n_K$ from Table
\ref{table2}, the analytic theory predicts that the apsidal and nodal
precessions are at nearly equal rates in opposite directions, with the
prograde apsidal precession having a period of $42.2$ years and the
retrograde nodal precession having a period of $42.8$ years.
These are in good agreement with the numerical results but slightly
shorter for the apsidal precession and longer for the nodal
precession.

\begin{figure}
\plotone{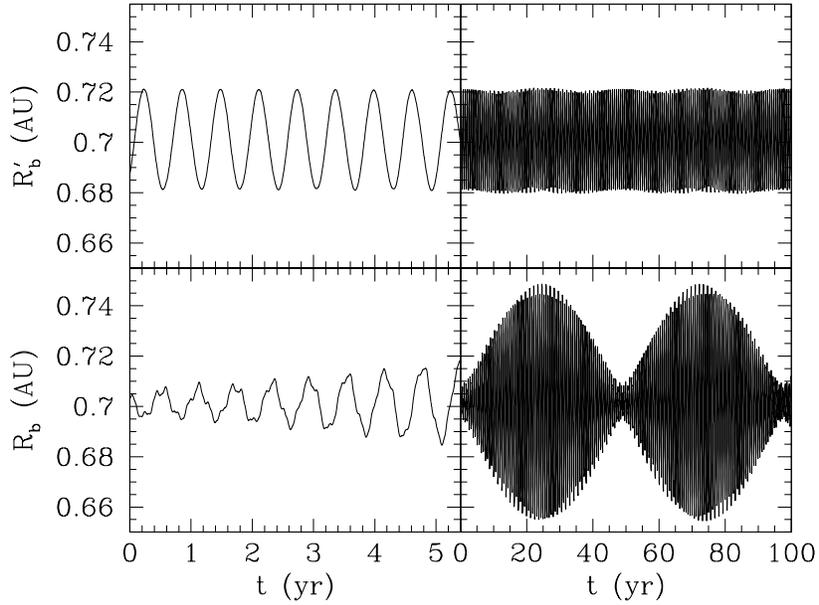}
\caption{
Variations in the orbital radius $R_b$ ({\it bottom panels}) and the
transformed orbital radius $R_b'$ ({\it top panels};
eq.~[\ref{Rprime}]) of the planet Kepler-16~b over several years
({\it left panels}) and over $100\,$yr ({\it right panels}) from
direct numerical orbit integration.
\label{fig:16radius}
}
\end{figure}

In the bottom panels of Figure \ref{fig:16radius} we plot the
variations in the orbital radius $R_b$ of the planet Kepler-16~b over
$5.4$ years and $100$ years.
Significant periodic variations in the amplitude of the oscillations
in $R_b$ are observed from the $100$-year plot.
The variations have a period of $48.6$ years, which is equal to the
period of apsidal precession.
The variations are the result of the superposition of the epicyclic
motion at frequency $\kappa_0$ with amplitude $e_{\rm free}$ and the
forced oscillation at frequency $n_0$ with amplitude $C_1^-$.
Without any loss of generality, we can assume that both $e_{\rm free}$
and $C_1^- \ge 0$.
After some algebraic manipulation using sum and product formulae of
trigonometric functions, these two terms in equation (\ref{Rt}) can be
written as
\begin{eqnarray}
& & e_{\rm free} \cos(\kappa_0 t + \psi) + C_1^- \cos(\phi_0 - \varpi_B)
\nonumber\\
& & \quad = e_{\rm free} \cos(\kappa_0 t + \psi)
    + C_1^- \cos(n_0 t + \varphi_0 - \varpi_B)
\nonumber\\
& & \quad = (C_1^-+e_{\rm free})
    \cos\left(\varpi_{\rm free} - \varpi_B \over 2\right)
    \cos\left[(n_0+\kappa_0) t + \varphi_0 - \varpi_B + \psi\over 2\right]
\nonumber\\
& & \qquad -(C_1^--e_{\rm free})
    \sin\left(\varpi_{\rm free} - \varpi_B \over 2\right)
    \sin\left[(n_0+\kappa_0) t + \varphi_0 - \varpi_B + \psi\over 2\right] ,
\label{env}
\end{eqnarray}
where $\varpi_{\rm free} = (n_0-\kappa_0) t + \varphi_0 - \psi$ is the
{\it free} longitude of periapse.
A maximum amplitude is reached when $\varpi_{\rm free} - \varpi_B  =
2\ell\pi$ (where $\ell$ is an integer), in which case the right-hand
side of equation (\ref{env}) becomes $\pm (C_1^-+e_{\rm free})
\cos\{[(n_0+\kappa_0) t + \varphi_0 - \varpi_B + \psi]/ 2\}$.
Similarly, a minimum amplitude is reached when $\varpi_{\rm free} -
\varpi_B = (2\ell + 1)\pi$, in which case the right-hand side of
equation (\ref{env}) becomes $\pm (C_1^--e_{\rm free})
\sin\{[(n_0+\kappa_0) t + \varphi_0 - \varpi_B + \psi]/2\}$.
Therefore, a maximum (or minimum) amplitude is reached every
$2\pi/|n_0-\kappa_0|$, which is equal to the apsidal precession
period.
The small minimum amplitude and large variations in the amplitude
observed in the 100-year plot indicate that $e_{\rm free} \sim C_1^-$.

The $5.4$-year plot in the lower left panel of Figure
\ref{fig:16radius} clearly shows an increase in the amplitude of the
oscillations in $R_b$ at the initial epoch due to the changing
relative phase between the free and forced eccentricity terms, as well
as higher-frequency forced oscillations.
In order to study in more detail the forced oscillations and epicyclic
motion, we plot in the upper panels of Figure \ref{fig:16radius} a
transformed orbital radius defined by
\begin{eqnarray}
R' &=& R + R_0 \Bigg\{C_0 \cos M_B + \sum \Big[C_k^0 \cos k(\phi_0-M_B-\varpi_B) \nonumber\\
& & + C_k^+ \cos (k(\phi_0-\varpi_B)-(k+1)M_B)
+ C_k^- \cos (k(\phi_0-\varpi_B)-(k-1)M_B) \Big]\Bigg\},
\label{Rprime}
\end{eqnarray}
with $R_0$, $C_0$, $C_k^0$, and $C_k^\pm$ from Table \ref{table2} and
$\phi_0$, $M_B$, and $\varpi_B$ from the numerical integration itself
(which eliminates phase errors due to small frequency errors and the
very slow precession of the binary's periapse).
It is clear from a comparison between the upper and lower panels of
Figure \ref{fig:16radius} that the forced oscillations (including the
forced eccentricity term) are sufficiently close to those predicted by
the analytic theory that they are effectively eliminated in $R_b'$.
The largest forced oscillation term, other than the forced
eccentricity term, is $C_2^-$ with a fractional amplitude of $0.0024$
and a period of $2\pi/(2 n_0 - n_{AB}) = 64.5$ days, and the other
forced oscillation terms are at least a factor of $4$ smaller in
amplitudes.
The free eccentricity can be easily obtained from $R_b'$, which shows
only periodic epicyclic variation at frequency $\kappa_0$.
The forced and free eccentricities of Kepler-16~b are $C_1^- = 0.036$
and $e_{\rm free} = 0.030$, respectively.

\begin{figure}
\plotone{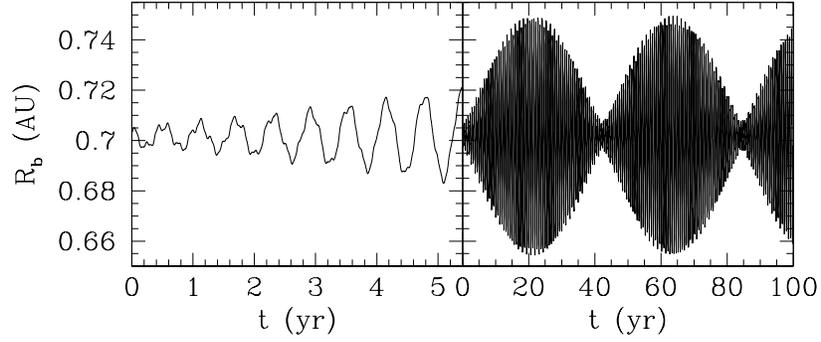}
\caption{
Variations in the orbital radius $R_b$ of the planet Kepler-16~b over
several years ({\it left panel}) and over $100\,$yr ({\it right
panel}) according to equation (\ref{Rt}) of the analytic theory with
$e_{\rm free} = 0.030$.
\label{fig:16anaR}
}
\end{figure}

\begin{figure}
\plotone{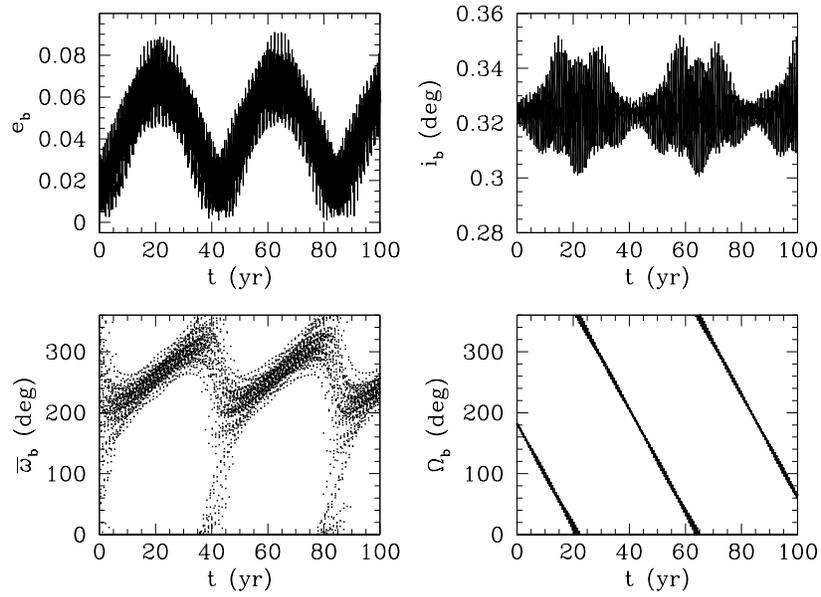}
\caption{
Same as Fig.~\ref{fig:16orbel}, but according to the analytic theory
with $e_{\rm free} = 0.030$.
\label{fig:16anal}
}
\end{figure}

Having obtained $e_{\rm free}$, which is a free parameter of the
analytic theory, we can now directly plot the evolution of $R$
according to equation (\ref{Rt}) of the analytic theory
(Fig.~\ref{fig:16anaR}), as well as the evolution of the osculating
Keplerian elements using $R$ from equation (\ref{Rt}), $\phi$ from
equation (\ref{phit}), and their time derivatives
(Fig.~\ref{fig:16anal}).
They are in excellent agreement with the numerical orbit integration
shown in Figures \ref{fig:16orbel} and \ref{fig:16radius}, except for
the faster periapse precession (and hence faster periodic variations
in the radial oscillation amplitude), the slower nodal precession, and
the lack of periodic variations in the inclination.
Our analytic theory only gives the free oscillations in the vertical
direction and cannot explain the periodic variations in the
inclination at twice the nodal precession rate.
For the eccentricity variations, if we ignore the higher-frequency
forced oscillations, equation (\ref{env}) shows that the osculating
eccentricity reaches a maximum of $C_1^- + e_{\rm free} = 0.066$ when
the free longitude of periapse $\varpi_{\rm free}$ is aligned with the
forced longitude of periapse (which is equal to the longitude of
periapse of the binary $\varpi_B$; see \S \ref{sec:solution}), and
reaches a minimum of $|C_1^- - e_{\rm free}| = 0.006$ when the free
longitude of periapse $\varpi_{\rm free}$ is anti-aligned with the
forced longitude of periapse.
This behavior agrees with the usual definitions of the forced and free
eccentricities and longitudes of periapse (see, e.g., Section 7.4 of
\citealt{mur99}).

\subsection{Kepler-34}
\label{sec:kepler34}

\begin{figure}
\plotone{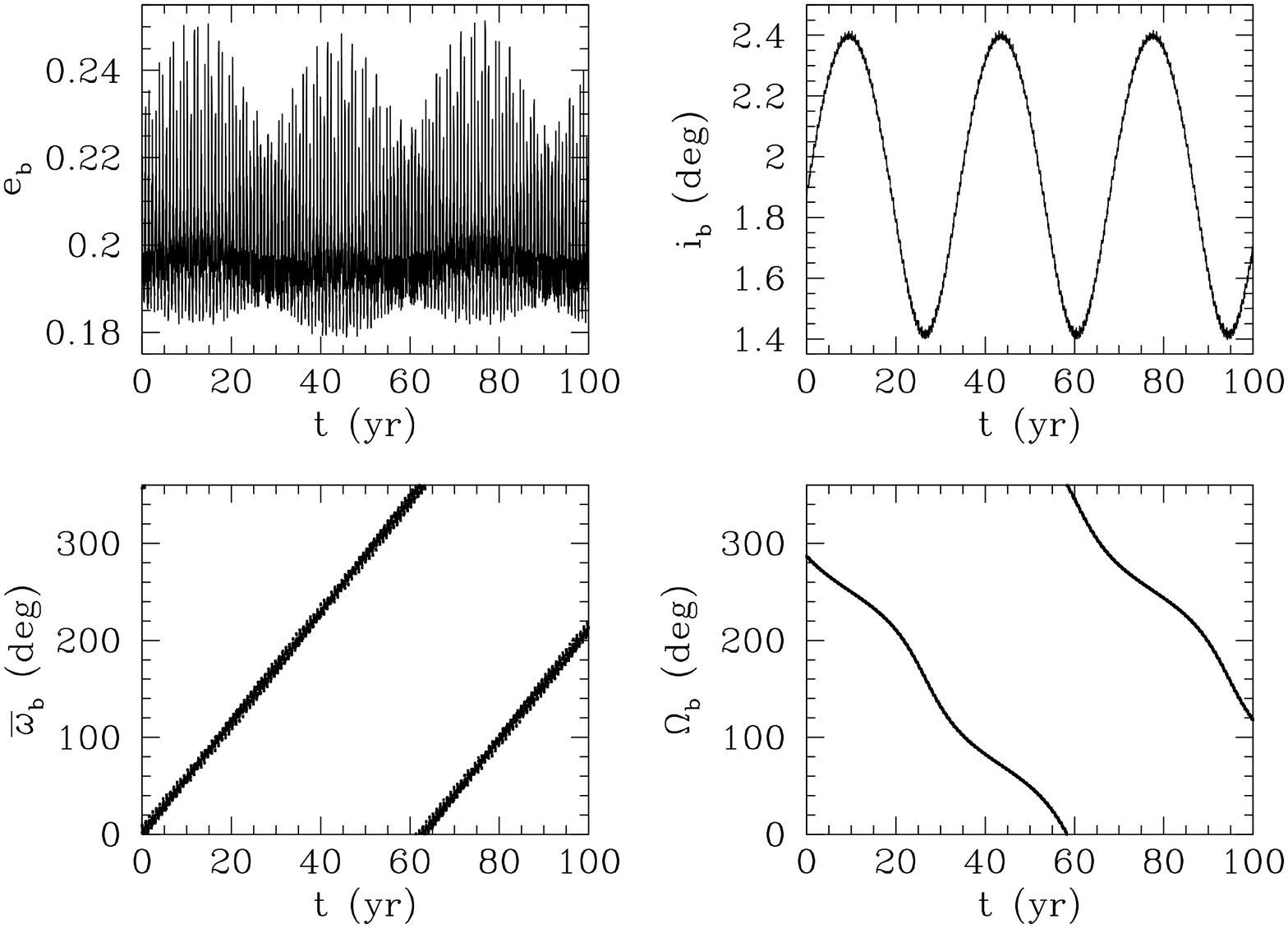}
\caption{
Same as Fig.~\ref{fig:16orbel}, but for the planet Kepler-34~b.
\label{fig:34orbel}
}
\end{figure}

\begin{figure}
\plotone{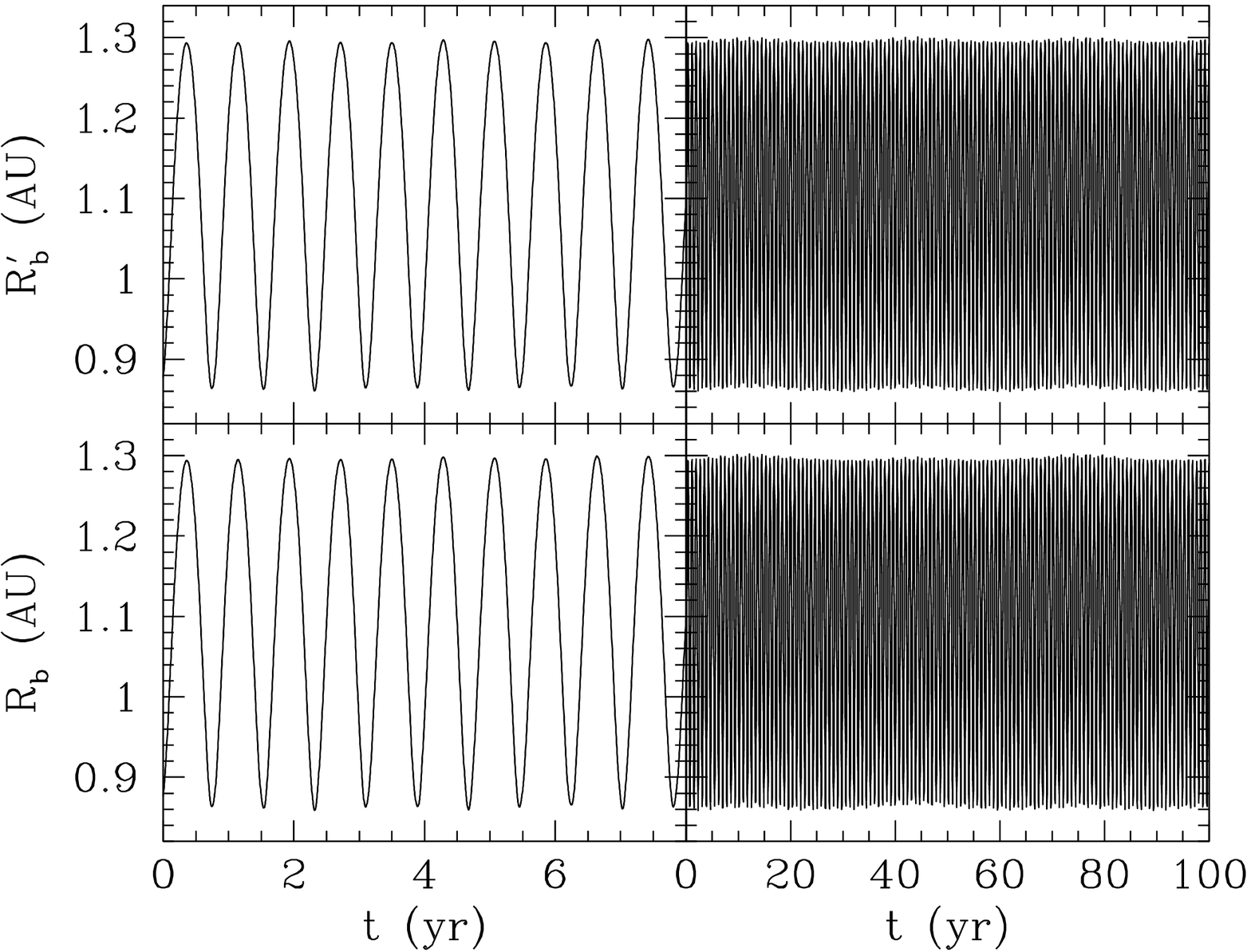}
\caption{
Same as Fig.~\ref{fig:16radius}, but for the planet Kepler-34~b.
\label{fig:34radius}
}
\end{figure}

Figures \ref{fig:34orbel} and \ref{fig:34radius} show the evolution of
the osculating Keplerian orbital elements, the orbital radius $R_b$
and the transformed orbital radius $R_b'$ of the planet Kepler-34~b.
The periods of prograde apsidal precession and retrograde nodal
precession are $62.9$ and $67.9$ years, respectively, from the
numerical orbit integration.
The analytic theory predicts $91.1$ and $91.9$ years, respectively,
which are longer than the numerical results by more than $35\%$.
The main reason for the large discrepancy is that the analytic theory
is only accurate to first order in the binary eccentricity $e_{AB}$
and Kepler-34 has the largest $e_{AB}$($= 0.52$) among the three
systems.
We shall derive in \S \ref{sec:discussion} simple corrections at the
second order in $e_{AB}$ that significantly improve the analytic
predictions of the precession periods.

One might think that the large osculating Keplerian eccentricity ($e_b
\sim 0.2$) is due to forcing by the eccentric binary.
But the nearly identical plots of $R_b$ and $R_b'$ show that the
variations in $R_b$ are primarily due to epicyclic motion and that the
forced eccentricity $C_1^-$ and other forced oscillation terms are
small.
Indeed, $C_1^- = 0.0019$, which is smaller than that for Kepler-16~b
by more than an order of magnitude, and the next largest forced
oscillation term is $C_2^- = 0.00068$ (see Table \ref{table2}).
The forced eccentricity $C_1^-$, as well as all $C_k^0$ and $C_k^\pm$
terms with odd $k$, are small because of the nearly equal masses of
the binary components of Kepler-34 ($m_B/m_A = 0.97$; see discussion
in the last paragraph of \S \ref{sec:solution}).
We find from the variations in $R_b'$ that the free eccentricity
$e_{\rm free} = 0.204$.

\subsection{Kepler-35}
\label{sec:kepler36}

\begin{figure}
\plotone{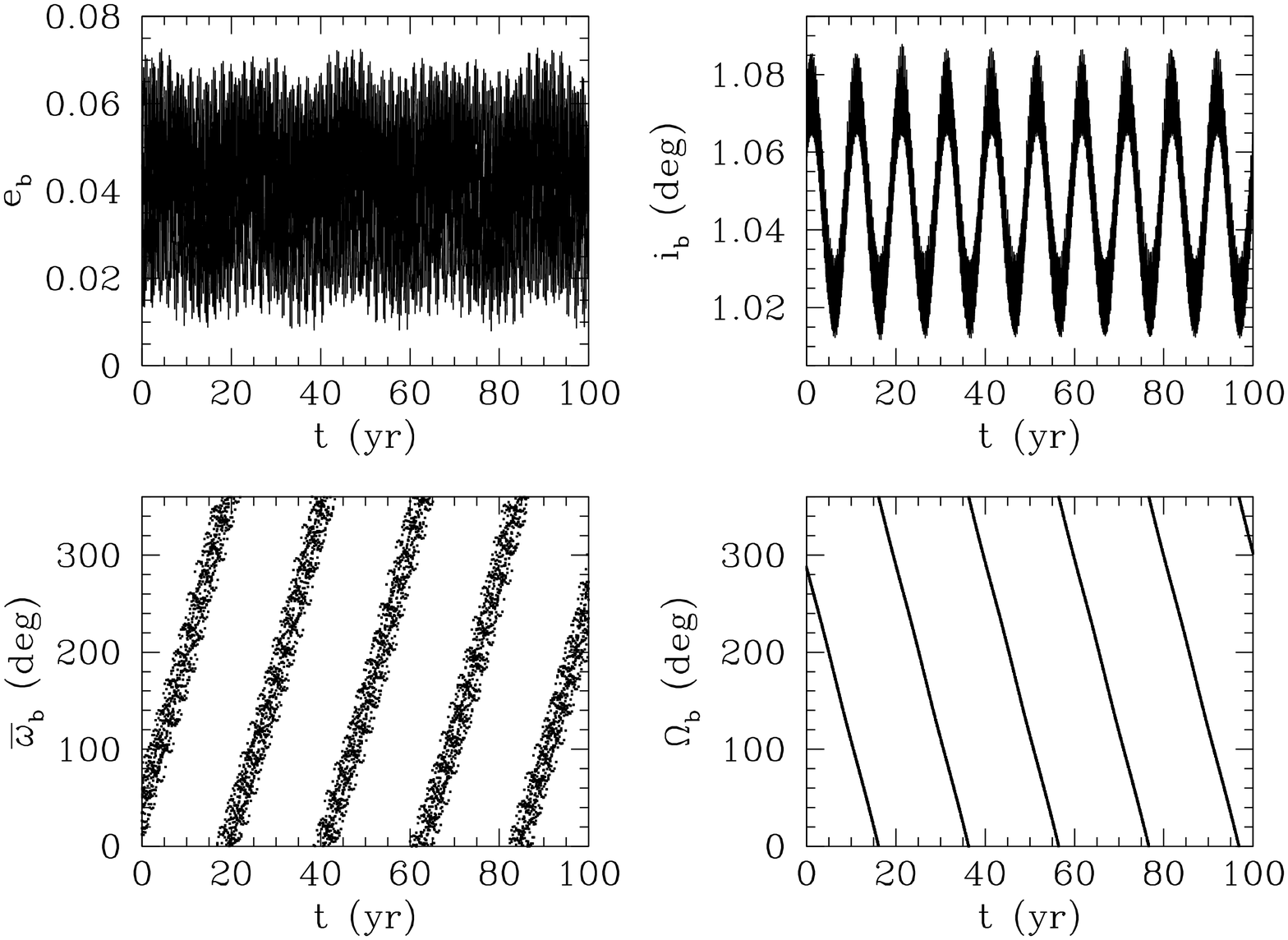}
\caption{
Same as Fig.~\ref{fig:16orbel}, but for the planet Kepler-35~b.
\label{fig:35orbel}
}
\end{figure}

For Kepler-35 which has binary eccentricity ($e_{AB} = 0.14$) similar
to Kepler-16, the numerical and analytic apsidal and nodal precession
periods of the planet are in good agreement.
The numerically determined apsidal and nodal precession periods are
$21.7$ and $20.2$ years, respectively (see Fig.~\ref{fig:35orbel}),
and the analytic ones are $20.4$ and $20.8$ years, respectively.

\begin{figure}
\plotone{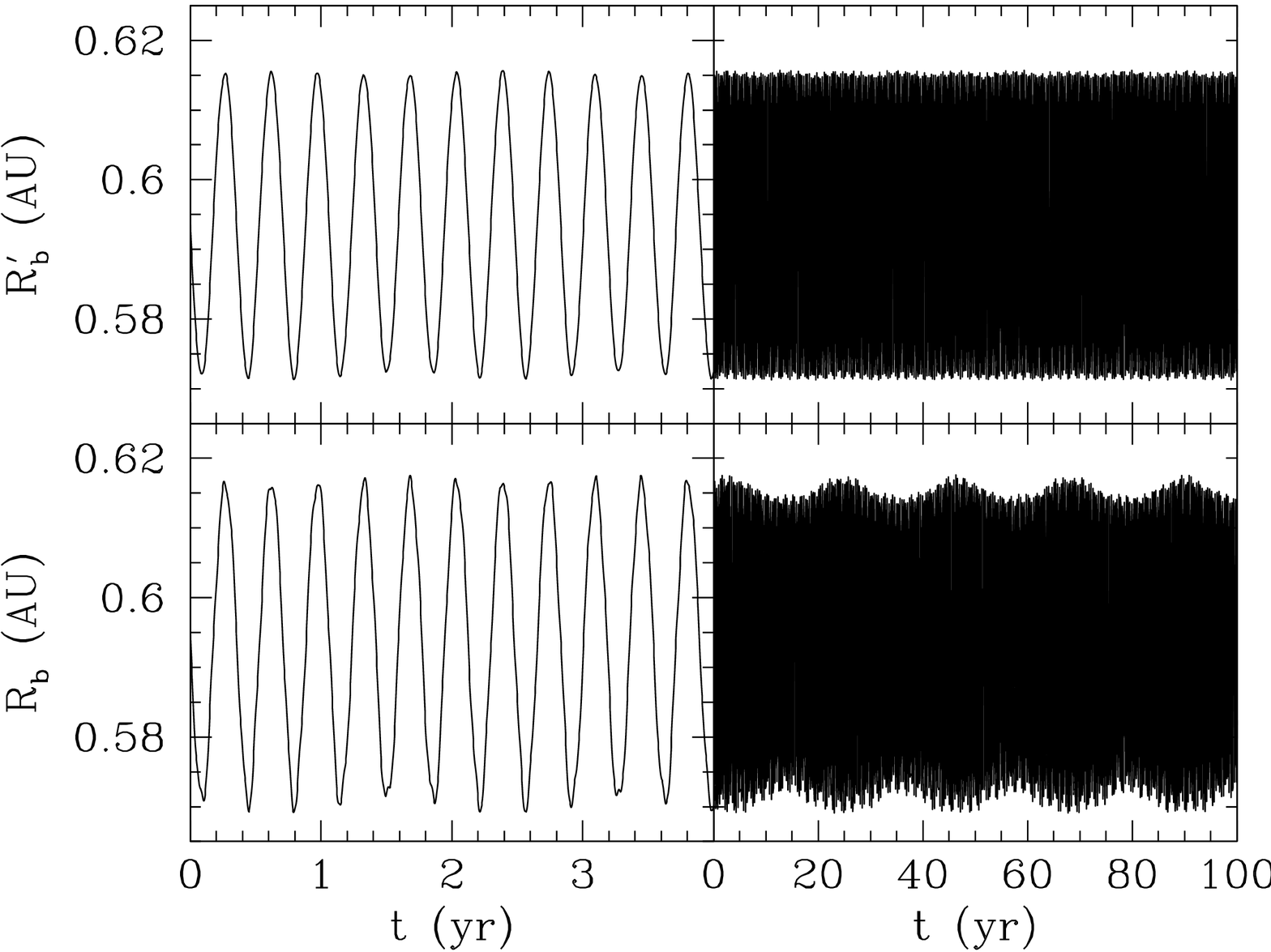}
\caption{
Same as Fig.~\ref{fig:16radius}, but for the planet Kepler-35~b.
\label{fig:35radius}
}
\end{figure}

As in the case of Kepler-34, the binary components have nearly equal
masses ($m_B/m_A = 0.91$) and the forced eccentricity of the planet
$C_1^- = 0.0025$ is small.
However, because the free eccentricity is much smaller than that for
Kepler-34~b and comparable to that for Kepler-16~b, moderate
variations in the amplitude of oscillations in $R_b$ with the same
period as the apsidal precession are clearly observed in the
$100$-year plot in the lower right panel of Figure \ref{fig:35radius}.
The $4$-year plot in the lower left panel of Figure \ref{fig:35radius}
also shows the effects of higher-frequency forced oscillations terms.
As for Kepler-16~b and 34~b, $C_2^-$ is the largest forced oscillation
term after the forced eccentricity term $C_1^-$ (see Table
\ref{table2}).
The forced oscillations are sufficiently close to those predicted by
the analytic theory that the transformed orbital radius $R_b'$ shows
only periodic epicyclic variation at frequency $\kappa_0$ (see upper
panels of Fig.~\ref{fig:35radius}).
The free eccentricity from the variations in $R_b'$ is $e_{\rm free} =
0.038$.

\section{DISCUSSION}
\label{sec:discussion}

The analytic theory developed in \S \ref{sec:theory} is accurate to
first order in the binary eccentricity $e_{AB}$ and to first order in
the deviations $R_1$, $\phi_1$, and $z_1$ of the planetary motion from
the circular motion of the guiding center.
It also treats the planet as a test particle and ignores the
gravitational effects of the planet on the motion of the binary.
From the comparisons with direct numerical orbit integrations of the
Kepler-16, 34, and 35 systems in \S \ref{sec:numerical}, we have shown
that the analytic theory gives an accurate description of the
planetary motion in all three cases, except for the apsidal and nodal
precession periods of Kepler-34~b with $e_{AB} = 0.52$.

It was pointed out in \S \ref{sec:theory} that the expressions for
the mean motion $n_0$ (eq.~[\ref{n0}]), the epicyclic frequency
$\kappa_0$ (eq.~[\ref{kappa0}]), and the vertical frequency $\nu_0$
(eq.~[\ref{nu0}]) involve the axisymmetric $\Phi_{000}$ and
$\Phi_{200}$ components of the potential and that the axisymmetric
components of the potential are identical to those due to two rings:
one of mass $m_A$ and radius $a_A$ and another of mass $m_B$ and
radius $a_B$.
If we expand $R_B/a_B$ to higher orders in $e_{AB}$ (see eq.~[2.81] of
\citealt{mur99}),
\begin{equation}
{R_B \over a_B}
= 1 - e_{AB} \cos M_B + {e_{AB}^2 \over 2} (1 - \cos 2 M_B)
  + {3 e_{AB}^3 \over 8} (\cos M_B - \cos 3 M_B) + \cdots ,
\end{equation}
which means that the time-averaged $R_B/a_B = 1 + e_{AB}^2/2$ and that
it might be more appropriate for the axisymmetric components of the
potential to be due to two rings of radii $a_A (1 + e_{AB}^2/2)$ and
$a_B (1 + e_{AB}^2/2)$.
This is achieved if we selectively include just the $e_{AB}^2/2$ term
beyond first order in $e_{AB}$ and use $R_B/a_B \approx 1 -
e_{AB} \cos M_B + e_{AB}^2/2$ (and similarly for $R_A/a_A$).
Then the only modifications to the analytic theory in \S \ref{sec:theory} 
are that $b_{(j+1)/2}^k(\alpha_{A})$ is replaced by
$b_{(j+1)/2}^k[\alpha_{A}(1+e_{AB}^2/2)]$, and
$b_{(j+1)/2}^k(\alpha_{B})$ by
$b_{(j+1)/2}^k[\alpha_{B}(1+e_{AB}^2/2)]$, in equation (\ref{Phijk0})
for $\Phi_{jk0}$, and similarly for $Db_{(j+1)/2}^k(\alpha_{A})$ and
$Db_{(j+1)/2}^k(\alpha_{B})$ in equation (\ref{Phijk1}) for
$\Phi_{jk1}$.
With this simple modification, the analytic predictions for the
apsidal and nodal precession rates are faster than the unmodified
values by only $2$--$3\%$ for Kepler-16~b and 35~b, but by $\sim 29\%$
for Kepler-34~b.
The increase by approximately $(1 + e_{AB}^2/2)^2$ can be understood
from the $(a_{AB}/R_0)^2$ scaling of the lowest order expression for
the precession rates in equation (\ref{preratesapprox}).
The modified analytic precession periods for Kepler-34~b are $71.4$
years for periapse and $72.1$ years for ascending node, which are much
closer to the numerical results ($62.9$ and $67.9$ years,
respectively) than the unmodified values ($\sim 91$ years).

The $1 + e_{AB}^2/2$ modification also affects the amplitudes of the
forced oscillation terms.
The change in the largest of these, the forced eccentricity $C_1^-$,
is small: $\la 6\%$ even for Kepler-34~b.
The change in the second largest forced oscillation term, $C_2^-$, is
approximately $1 + 5 e_{AB}^2/6$, which is only $2\%$ for Kepler-16~b
and 35~b but $\sim 24\%$ for Kepler-34~b.
However, as we saw in \S \ref{sec:kepler34}, both $C_1^-$ and $C_2^-$
are small compared to the free eccentricity and have no noticeable
effect on the evolution of $R_b$ for Kepler-34~b.

The $1 + e_{AB}^2/2$ modification that we have just described is not
rigorously correct.
We have attempted to construct an analytic theory that is accurate to
$O(e_{AB}^2)$.
Preliminary analysis indicates that there are no additional
corrections at $O(e_{AB}^2)$ for $C_2^-$ but that there are additional
corrections to the precession periods and $C_1^-$.
The full $O(e_{AB}^2)$ corrections to the precession periods and
$C_1^-$ of Kepler-16~b and Kepler-35~b, as well as $C_1^-$ of
Kepler-34~b, remain small (less than a few \%).
For Kepler-35~b, the apsidal (nodal) precession period may decrease
(increase) by a few years beyond the $1 + e_{AB}^2/2$ modification
discussed above.
The full $O(e_{AB}^2)$ theory also introduces new forced oscillation
terms with frequencies $|kn_0-(k\pm 2)n_{AB}|$, but they are likely
small in amplitudes, as they are not observed in the direct numerical
orbit integrations.

The most obvious effects of the gravitational force of the planet on
the binary are the precession of the binary's periapse and ascending
node.
Due to our choice of the invariable plane as the reference plane, the
longitude of ascending node of the binary must be $180^\circ$ from,
and precesses at the same rate as, the longitude of ascending node of
the planet.
From the direct numerical orbit integrations, we find that the
apsidal precession rates of the binary are $0.026$, $0.0033$ and
$0.0086$ degrees per year for Kepler-16, 34, and 35, respectively,
which are much smaller than those of the planet.
For the comparisons in \S \ref{sec:numerical}, this very slow
precession of the binary's periapse is taken into account by using
$\varpi_B$ from the numerical integrations when we plot $R'$ defined
in equation (\ref{Rprime}).

\section{CONCLUSIONS}
\label{sec:summary}

We have developed an analytic theory to model the motion of the
recently discovered circumbinary planets: Kepler-16~b, Kepler-34~b,
and Kepler-35~b.
Their orbits are significantly non-Keplerian due to the large
secondary-to-primary mass ratio and orbital eccentricity of the
binaries, as well as the proximity of the planets to the binaries.
The analytic theory in \S \ref{sec:theory} shows that the motion in
$R$ and $\phi$ can be represented by the superposition of the circular
motion of a guiding center at mean motion $n_0$, the epicyclic motion,
and the forced oscillations, and that the motion in $z$ decouples from
that in $R$ and $\phi$ and has only free oscillations.
One of the forced oscillation terms has frequency $n_0$ and can be
identified as the forced eccentricity, while the epicyclic motion at
frequency $\kappa_0$ can be identified as the free eccentricity.

Comparisons with direct numerical orbit integrations in \S
\ref{sec:numerical} show that the analytic theory (with the simple
modification in \S \ref{sec:discussion}) gives an accurate description
of the planetary motion for all three Kepler systems, including the
precession of the periapse and ascending node.
The analytic theory explains the periodic variations in the amplitude
of oscillations of the orbital radius (which is most obvious for
Kepler-16~b and negligible for Kepler-34~b) by the superposition of
the epicyclic motion and the forced eccentricity oscillation.
The amplitude (and osculating eccentricity) variations have a period
equal to that of the apsidal precession as predicted by the theory.

The amplitude, $C_1^-$, of the forced eccentricity term is an explicit
function of the parameters of the binary and of the guiding center
radius of the planet in the analytic theory.
For Kepler-16~b, 34~b, and 35~b, $C_1^- = 0.036$, $0.0019$, and
$0.0025$, respectively.
The free eccentricity, $e_{\rm free}$, of the epicyclic motion is a
free parameter in the analytic theory and can be obtained from, e.g.,
the variations in the orbital radius of the planet.
For Kepler-16~b, 34~b, and 35~b, $e_{\rm free} = 0.030$, $0.204$, and
$0.038$, respectively.
Note that the Kepler-34 system with the largest binary eccentricity
($e_{AB} = 0.52$) has the largest $e_{\rm free}$ while the other two
Kepler systems with comparable $e_{AB}$ have comparable $e_{\rm free}$.
Since the free eccentricity is a free parameter that was set by
dynamical processes during the formation and/or subsequent evolution
of the circumbinary planet, the free eccentricity of the three Kepler
circumbinary planets (and additional circumbinary planets in the
future) should provide important clues to these processes.

While this paper was under review, three more circumbinary planetary
systems were announced: Kepler-38 and PH1 with one planet each and
Kepler-47 with two planets \citep{oro12a,oro12b,sch12}.
Direct numerical integrations and our analytic theory show that (i)
Kepler-38~b is similar to Kepler-16~b in having $e_{\rm free} \sim
C_1^- \sim 0.024$ and large variations in the amplitude of
oscillations in $R$; (ii) Kepler-47~b is similar to Kepler-35~b in
having $e_{\rm free}$ larger than $C_1^- \sim 0.004$ and moderate
variations in the amplitude of oscillations in $R$; (iii) Kepler-47~c
is similar to Kepler-34 in having $e_{\rm free}$ much larger than
$C_1^- \sim 0.001$ and negligible variations in the amplitude of
oscillations in $R$; and (iv) PH1 has $e_{\rm free} \sim 0.1$ and
$C_1^- \sim 0.04$.

\acknowledgments
It is a pleasure to thank Josh Carter for furnishing the best-fit
orbital parameters of the Kepler circumbinary planetary systems,
Daniel Fabrycky for informative discussions, and the referee for
helpful comments on the manuscript.
C.~K.~L.\ also thanks K.~H.\ Chan, X.\ Tan, and W.~Y.\ Li for
enlightening discussion.
This work was supported in part by Hong Kong RGC grant HKU 7034/09P.

% \clearpage

\clearpage

\begin{deluxetable}{lrrr}
\tablecolumns{4}
\tablewidth{0pt}
\tablecaption{Orbital Parameters of Circumbinary Planetary Systems
\label{table1}}
\tablehead{
\colhead{Parameter} 		& \colhead{Kepler-16} & \colhead{Kepler-34} & \colhead{Kepler-35}
}
\startdata
Epoch (BJD) & 2,455,000.0 & 2,454,900.0 & 2,454,900.0 \\
$G M_A$ ($10^{-4}\,$AU$^3\,$d$^{-2}$) & 2.0328 & 3.1045 & 2.6187 \\
$G M_B$ ($10^{-4}\,$AU$^3\,$d$^{-2}$) & 0.5987 & 3.0232 & 2.3903 \\
$G M_b$ ($10^{-8}\,$AU$^3\,$d$^{-2}$) & 9.3119 & 6.5822 & 3.6839 \\
\\
\multicolumn{4}{c}{Binary star orbit} \\
\\
Semimajor axis (AU)	& 0.22405 & 0.22847 & 0.17603 \\
Eccentricity 		& 0.16048 & 0.52068 & 0.14224\\
Inclination (deg)		& 0.0011 & 0.0020 & 0.0006 \\
Argument of periapse (deg)	& 257.79 & 323.86 & 338.95 \\
Long.\ ascending node (deg)	& 5.70 & 107.45 & 107.56 \\
Mean anomaly (deg)		& 129.84 & 52.66 & 299.31 \\
\\
\multicolumn{4}{c}{Planet orbit} \\
\\
Semimajor axis (AU)		& 0.72042 & 1.08617 & 0.60497 \\
Eccentricity			& 0.02373 & 0.20861 & 0.04845 \\
Inclination (deg)		& 0.3083 & 1.8590 & 1.0714 \\
Argument of periapse (deg)	& 304.05 & 69.41 & 91.17 \\
Long.\ ascending node (deg)	& 185.70 & 287.45 & 287.56 \\
Mean anomaly (deg)		& 358.85 & 17.75 & 292.17 \\

\enddata
\tablecomments{The orbital parameters are the best-fit osculating
Jacobian parameters relative to the invariable plane at the listed
epoch.}
\end{deluxetable}

\clearpage

\begin{deluxetable}{lrrr}
\tablecolumns{2}
\tablewidth{0pt}
\tablecaption{Parameters of Analytic Theory
\label{table2}}
\tablehead{
\colhead{Parameter} & \colhead{Kepler-16} & \colhead{Kepler-34} & \colhead{Kepler-35}
}
\startdata
$R_0$ (AU)        & $0.7016$ &    $1.0804$ &           $0.5933$ \\
$n_K$ (yr$^{-1}$) & $10.0823$ &   $8.0512$ &           $17.8875$ \\
$n_0/n_K$         & $1.00702$ &   $1.00423$ &          $1.00838$ \\
$\kappa_0/n_K$    & $0.99224$ &   $0.99567$ &          $0.99119$ \\
$\nu_0/n_K$       & $1.02158$ &   $1.01272$ &          $1.02527$ \\
$C_0$	          & $0.000159$ & $0.000085$ &        $0.000131$ \\
$C_1^0$           & $-0.000282$ &  $-6\times 10^{-7}$ &  $-0.000020$ \\
$C_2^0$           & $-0.000589$ &  $-0.000079$ &         $-0.000533$ \\
$C_3^0$           & $-0.000049$ &  $-1\times 10^{-7}$ &  $-0.000003$ \\
$C_1^+$           & $0.000005$ & $4\times 10^{-8}$ & $3\times 10^{-7}$ \\
$C_2^+$           & $-0.000033$ &  $-0.000016$ &         $-0.000028$ \\
$C_3^+$           & $-0.000006$ &  $-4\times 10^{-8}$ &  $-4\times 10^{-7}$ \\
$C_1^-$           & $0.035772$ & $0.001861$ &        $0.002493$ \\
$C_2^-$           & $0.002438$ & $0.000683$ &        $0.001731$ \\
$C_3^-$           & $0.000110$ & $7\times 10^{-7}$ & $0.000007$ \\
\enddata
\end{deluxetable}

\clearpage

% \clearpage

% \clearpage

% \clearpage

% \clearpage

% \clearpage

% \clearpage

% \clearpage

\begin{thebibliography}{}
\bibitem[Doyle et al.(2011)]{doy11}
     Doyle, L. R., et al. 2012, Science, 333, 1602
\bibitem[Goldreich \& Tremaine(1980)]{gol80}
     Goldreich, P., \& Tremaine, S. 1980, \apj, 241, 425
\bibitem[Lee \& Peale(2003)]{lee03}
     Lee, M. H., \& Peale, S. J. 2003, \apj, 592, 1201
\bibitem[Lee \& Peale(2006)]{lee06}
     Lee, M. H., \& Peale, S. J. 2006, Icarus, 184, 573
\bibitem[Murray \& Dermott(1999)]{mur99}
     Murray, C. D., \& Dermott, S. F. 1999, Solar System Dynamics
     (Cambridge: Cambridge Univ. Press)
\bibitem[Orosz et al.(2012a)]{oro12a}
     Orosz, J. A., et al. 2012a, Science, 337, 1511
\bibitem[Orosz et al.(2012b)]{oro12b}
     Orosz, J. A., et al. 2012b, ApJ, 758, 87
\bibitem[Schwamb et al.(2012)]{sch12}
     Schwamb, M. E., et al. 2012, ApJ, submitted (arXiv:1210.3612)
\bibitem[Welsh et al.(2012)]{wel12}
     Welsh, W. F., et al. 2012, Nature, 481, 475
\bibitem[Wisdom \& Holman(1991)]{wis91}
     Wisdom, J., \& Holman, M. 1991, \aj, 102, 1528
\end{thebibliography}
\end{document}